\documentclass[aps,12pt,
english,preprintnumbers,nofootinbib,
]{revtex4-1}

\usepackage{amsfonts,amsmath,amssymb}
\usepackage{graphicx}
\usepackage[utf8]{inputenc}
\usepackage{hyperref}
\usepackage{babel}
\usepackage{enumitem}
\usepackage{wrapfig}

\newcommand\e{{\rm e}}

\newcommand\be{\begin{equation}}
\newcommand\ee{\end{equation}}
\newcommand\bea{\begin{eqnarray}}
\newcommand\eea{\end{eqnarray}}

\begin{document}

\def\rhoo{\rho_{_0}\!} 
\def\rhooo{\rho_{_{0,0}}\!} 

\begin{flushright}
\phantom{
{\tt arXiv:2006.$\_\_\_\_$}
}
\end{flushright}


\title{Wigner meets 't Hooft near the black hole horizon}
\author{Clifford V. Johnson}
\affiliation{\medskip\\ Department of Physics, 
Princeton University, Princeton, NJ 08544-0708, U.S.A.\vskip0.15cm}
 \affiliation{\medskip\\ Department of Physics and Astronomy, University of
Southern California,
 Los Angeles, CA 90089-0484, U.S.A.}


\begin{abstract}
\centerline{\tt cvj@princeton.edu}
\bigskip
\bigskip
\bigskip

Recent work on Euclidean quantum gravity, black hole thermodynamics, and the holographic principle has seen the return of random matrix models as a powerful tool. It is explained how they allow for the study of the physics well beyond the  perturbative expansion. In fact, a fully non-perturbative treatment naturally unites the familiar approach of summing over smooth geometries of all topologies  with the statistical approach to characterizing the typical properties of a  Hamiltonian. Remarkably, this leads to an explicit excavation of the underlying microstates of quantum gravity that has applications to the low temperature dynamics of a large class of black holes.
\vskip1.5cm
\centerline{30th March 2022 (Tiny corrections:  7th June 2022)}
\vskip1.5cm
\centerline{\it Essay written for the Gravity Research Foundation 2022 Awards for Essays on Gravitation.}

\end{abstract}


\maketitle

{\it Background}---The Euclidean approach to quantum gravity~\cite{Hartle:1976tp,Gibbons:1976ue}, while very powerful,  contains many puzzles. Among these is the issue of what the rules are for including contributions to the partition function. The intuition to sum over all geometries {\it and} all topologies of spacetime seems natural, but it is not clear if this is generally correct, and moreover it is technically difficult to perform such a sum in most cases.  Crucially, interpreting the results of such a computation is difficult, not the least because there is no guidance from experiment. 

Some strong guidance has come from the holographic principle, first proposed by 't~Hooft~\cite{tHooft:1993dmi}, which relates a theory of gravity in $D$ dimensions to a {\it non-gravitational} theory in $D{-}1$ dimensions. The latter theory, referred to here as the ``holographic dual'', can often be regarded as residing on the boundary of spacetime. Results of computations in the quantum gravitational theory can then be made sense of entirely in terms of the 
non-gravitational theory where the ordinary rules of quantum theory apply. Examples of such an holographic setup first appeared in the context of string theory~\cite{Susskind:1994vu}, most concretely as the AdS/CFT correspondence, constructed by Maldacena~\cite{Maldacena:1997re} and swiftly expanded upon by Gubser, Klebanov and Polyakov~\cite{Gubser:1998bc}  and Witten~\cite{Witten:1998qj,Witten:1998zw}.
The gravity  side has spacetimes that are asymptotically anti-de Sitter (AdS), while the holographic duals are conformal field theories (CFTs), some of which are very familiar, such as ${\cal N}{=}4$ supersymmetric $SU(N)$ Yang-Mills theory in 4D, where $N$ is large. 

In particular, Witten showed~\cite{Witten:1998zw} that because of the duality, the Hawking-Page phase transition~\cite{Hawking:1982dh} in  Euclidean quantum gravity gets a precise precise meaning in the dual Yang-Mills theory. In the  $D{=}5$ example the 4D spacetime boundary  is $S^1_\beta\,{\times}\, S^{3}$, where the first factor is the circle of Euclidean time $\tau$ with period $\beta{=}1/T$. The intuitive Euclidean approach is to integrate over all possible geometries, and sum over all possible topologies, of bulk spacetimes that have this $S^1_\beta\,{\times}\, S^{3}$ boundary. There are two solutions that can be considered as leading saddles of the action, one is (thermal) AdS$_5$ itself and the other is the AdS$_5$-Schwarzschild black hole. These two solutions exchange dominance in a first order phase transition as the  temperature $T_c$ is crossed. This represents the dual Yang-Mills theory's confinement/deconfinement phase transition: Above $T_c$, the  ${\sim}N^2$ degrees of freedom of the theory become manifest in a fluid/gas phase, as explicitly evident in the Bekenstein-Hawking~\cite{Bekenstein:1973ur,Hawking:1974sw}  entropy of the dual black hole solution: $S_{\rm BH}{=}A/G^{(5)}_{\rm N}{\sim}N^{2}$. 

This gives a strong indication that at least in this holographic context, summing over topological sectors makes sense, and would seem to be essential. However, this example, although embedded in a complete theory, is still a semi-classical piece of a more involved story. One might ask about what the calculus of non-saddle geometries looks like, and what the resulting physics is.  
To address such a question, given the limitations of techniques for integration and summation over all geometries and topologies in higher dimensions, it is useful to go to lower dimensions. The case of two dimensions is particularly tractable, since smooth Euclidean 2D surfaces  are simple to classify. 
The enumeration of all geometries can be done by temporarily discretizing them using polygons and then taking a continuum limit to extract the physics. This was appreciated long ago in the 1980s (see {\it e.g.,} refs.~\cite{David:1984tx,Kazakov:1985ea,David:1985et}), and a dramatic advance was made when it was discovered that such an enumeration process,  as well as summing over all topologies, was contained in the ``double scaling limit'' of  random matrix models~\cite{Gross:1990vs,Brezin:1990rb,Douglas:1990ve,Gross:1990aw}. 

{\it Enter 't Hooft}---The heart of the  technique goes  back to 't Hooft's  observations about the~$1/N$ expansion~\cite{'tHooft:1973jz}. $N{\times}N$ matrix valued propagating fields (such as gluons in an $SU(N)$ gauge theory)  have Feynman diagrams with double lines or ribbons, each edge present for the consistent propagation of each of the matrix indices. A  diagram comes with a definite power of $N$ 
 corresponding to the topology of the 2D surface upon which the Feynman diagram can be drawn without the ribbons crossing each other. For oriented vacuum diagrams, order $N^2$  is carried by all diagrams that can be drawn on the sphere,  order $N^0$ is the torus, $N^{-2}$ the double torus, and so forth. Generally the factor is $N^\chi$ where $\chi$ is the Euler number.  Placing a dot into each loop of  a diagram and connecting them by edges (dual to  propagators) yields a tessellation of the surface into polygons. For example a potential $V(M){=}\frac12M^2+g_4M^4$  gives a tessellation made  squares. The  partition sum of a model of just pure matrices~\cite{Brezin:1978sv,Bessis:1980ss}:
\be
\label{eq:partfun}
{\tilde Z} = \int dM \exp(-N{\rm Tr}[V(M)])\ ,
\ee
evaluated by all possible diagrams is a sum over all  tessellations, and of all possible topologies, organized perturbatively by the $1/N$ expansion about large $N$.   The ``double scaling'' continuum limit  takes  $N{\to}\infty$ while also tuning the coupling  $g_4$ to  a critical value~$g_{4}^{c}$ where surfaces with large numbers of squares  dominate the partition sum: In physical area units, the surfaces are now smooth, built from an infinite number of infinitesimally small squares. A scaled piece of the~$1/N$ expansion parameter, usually denoted $\hbar$ (but it is not Planck's constant!),  remains finite in this limit, and  its powers count the topology of  the diagrams according to $\hbar^{-\chi}$. More general polynomials $V(M){=}\frac12M^2+ \sum_p g_p M^p $ allow more couplings to be tuned to give a large class of 2D gravity models. The original context  (and also later  during a resurgence of interest in such models in the early 2000s, see {\it e.g.,}~refs.~\cite{McGreevy:2003kb,Klebanov:2003km,McGreevy:2003ep,Martinec:2003ka,Alexandrov:2003nn,Schomerus:2003vv,Sen:2003iv,Klebanov:2003wg,Maldacena:2004sn}) was the sum over Euclidean surfaces needed for  string theory's world-sheet path integral (then~$\hbar$ is the closed string coupling), and  so many of the foci were different from those we began with here. 
As a result, several instructive features of the matrix model as a tool for quantum gravity either went entirely overlooked, or simply under-appreciated, until very recently. 

{\it JT Gravity, leading order}---A special kind of 2D gravity theory,  discovered in the 1980s by Jackiw~\cite{Jackiw:1984je} and Teitelboim~\cite{Teitelboim:1983ux}, has been shown to appear very naturally as a universal low energy sector of many kinds of gravitational theories~\cite{Achucarro:1993fd,Nayak:2018qej,Ghosh:2019rcj,Kolekar:2018sba}. A simple example is the  case of near-extremal ({\it i.e.,} low temperature) black holes in $D{=}4$ gravity. Importantly, this  {\it can} be gravity in an asymptotically flat universe such as ours. Hence, this is especially interesting since progress in understanding the physics of this sector  will give direct insights into potential  quantum gravity phenomena in our universe, even though the gravity dynamics appears only effectively 2D. This is because near the horizon of  extremal ($T{=}0$)  black holes, the spacetime factorizes into an $(r,t)$ piece which is AdS$_2$, and an angular piece, an $S^2$ of fixed size~\cite{Bertotti:1959pf,Robinson:1959ev}. The inverse of the cosmological constant of the $(r,t)$ piece and the squared radius of the 
$S^2$ are given by~$Q^2$, in the case of the charge $Q$ Reissner-Nordstrom solution. The  Euclidean calculus gives the Bekenstein-Hawking entropy at extremality, denoted $S_0$, as 
$S_0{=}\pi Q^2$. 

Moving  to the near-extremal case ($T\,{\neq}\,0$ but small)  the dynamics is described by JT gravity, which captures the deviation away from pure AdS$_2$ with fixed $S^2$. Its action can be obtained by dimensional reduction on the $S^2$ from the   4D Einstein-Maxwell system to 2D.
 In  Euclidean signature it is,  on a 2D manifold ${\cal M}$ (with boundary $\partial{\cal M}$): 
\begin{eqnarray}
\label{eq:JT-gravity-action}
I = - \frac12\int_{\cal M}\!\!\sqrt{g} \phi(R+2) -\int_{\partial \cal  M}\!\!\sqrt{h} \phi_b (K-1) 
-\frac{S_0}{2\pi}\left(\frac12\int_{\cal M} \!\!\sqrt{g} R +\int_{\partial {\cal M}}\sqrt{h}K\right)\ , 
\end{eqnarray}
where $R$ is the Ricci scalar of 2D metric $g_{ij}$,  and  in the  boundary terms,~$K$ is the trace of the extrinsic curvature for induced metric $h_{ij}$. The scalar  $\phi$ captures the deviation of the $S^2$ away from area $4S_0$ and $\phi_b$ is its boundary value. Its linear coupling means that  integrating it out yields the local AdS$_2$ condition $R\,{=}{-}2$ and leaves only dynamics on the fluctuating boundary, which has total length~$\beta{=}1/T$. This is the period of Euclidean time $\tau=it$. The dynamics is given by the ``Schwarzian'' action~\cite{Almheiri:2014cka,Maldacena:2016hyu,Jensen:2016pah,Maldacena:2016upp,Engelsoy:2016xyb}. So far, ${\cal M}$ has  disc topology. 
 
Quantizing the Schwarzian dynamics yields an exact~\cite{Stanford:2017thb} (in $\beta$) result for the partition function $Z_0(\beta)$, which can be written as the  Laplace transform of a spectral density $\rhoo(E)$:  
\be
\label{eq:leading-relation}
Z_0(\beta) = \e^{S_0}\frac{\e^{\frac{\pi^2}{\beta}}}{4\sqrt{\pi}\beta^{\frac32}} = \int \rhoo(E)\e^{-\beta E}dE\ ,\qquad \rhoo(E) = \e^{S_0}\frac{\sinh{(2\pi\sqrt{E})}}{4\pi^2}\ .
\ee
There are in principle corrections to this result  away from the large $S_0$ regime.
It is natural to ask what physics such corrections should capture. Recall that large $S_0$ is the regime where semi-classical quantum gravity computations for the parent black hole are reliable. This gave, for example, the Bekenstein-Hawking entropy~$S_0$, and the Schwarzian physics has already captured the leading finite $T$ corrections to that: $S{=}S_0+2\pi^2T +\cdots$. What other physics should be sought?  A finite entropy indicates that the Hilbert space of the quantum gravity system is finite. Hence, a discrete structure should show up in the available energy spectrum, with spacing set by ${\rm e}^{-S_0}$. This is invisible in the large $S_0$ approximation being worked in so far, and so it is natural that the Schwarzian spectrum is continuous.

Rather elegantly~\cite{Saad:2019lba}, perturbation theory about the leading result,  organized in small $\e^{-S_0}$,  can be interpreted as a topological expansion because  in the action~(\ref{eq:JT-gravity-action}), $S_0$ multiplies $\chi{=}2{-}2g{-}b$ (the Euler number of the  $(r,\tau)$ manifold ${\cal M}$; $g$ counts handles and $b$ boundaries). So the partition function (and hence the spectral density) should be expected to be written: 
\begin{equation}
\label{eq:partition-sum}
Z(\beta)=\sum_{g=0}^\infty Z_g(\beta)+\cdots = \int\! \rho(E) \e^{-\beta E}dE\ , \qquad \rho(E)=\sum_{g=0}^\infty \rho_g(E)+\cdots
\end{equation} 
where $Z_g(\beta)$ is the contribution from   manifolds ${\cal M}$ with Euler number $\chi$ but with $b{=}1$ for the boundary of length $\beta$. It contains a factor~$\hbar^{2g-1}$, where for short, $\hbar{=}\e^{-{S_0}}$.  The ellipsis denote non-perturbative contributions.

Notice that the same symbol $\hbar$ is used here as was used earlier for the ``renormalized $1/N$'' topological expansion parameter that arose in the double-scaled matrix model approach to 2D Euclidean sums, and there is good reason for that.  A landmark paper  by  Saad, Shenker and Stanford~\cite{Saad:2019lba}  showed (aided by using powerful mathematical results of Mirzakhani~\cite{Mirzakhani:2006fta}, and of Eynard and Orantin~\cite{Eynard:2007fi} for the moduli space of hyperbolic surfaces) that the entire series of  perturbative corrections is captured by a random matrix model! 

It is worth pausing to admire the outcome so far. Rather than studying just  the contribution of the dominant saddles  give by the  two distinct topologies mentioned earlier in the 5D case (or just the AdS$_5$-Schwarzschild saddle alone if the boundary was $S^1_\beta{\times}\mathbb{R}^3$) the matrix model does the full path integral over all spacetime geometries and topologies. This,  as already remarked, is difficult in higher dimensions.

Having said that, it is very important to notice  that a continuous spectrum persists to all orders in the $\e^{-{S_0}}$ expansion. Each order in an expansion about $\rhoo(E)$ brings in another smooth function of~$E$, correcting the previous order, but not changing the fundamental fact that $\rho(E)$ is a smooth function.  On reflection this had to be the case given the starting point: perturbing away from the smooth leading result in some controlled manner is unlikely to yield a sensible discrete answer.  To see the underlying discrete structure promised by the Bekenstein-Hawking entropy requires a non-perturbative treatment of the matrix model. 

{\it Non-Perturbative Matrix Models}---Non-perturbative treatments of the matrix model from which  the perturbation theory emerged  has a history going back to the birth of the double scaling limit~\cite{Gross:1990vs,Brezin:1990rb,Douglas:1990ve,Gross:1990aw}. In fact, a great deal of excitement about the limit stemmed from the fact that it captured the physics of not just the all orders topological expansion, but physics beyond that. The manner in which this came about  was rather elegant. The  key quantity from which everything can be derived was encoded in a non-linear differential equation. Its asymptotic series  solution  yielded the topological expansion, but the full solution gave non-perturbative physics invisible at any order in the expansion. 

The nature of this non-perturbative physics in the string theory context was notoriously subtle to grapple with. An important early clue was that it had to do with the behaviour of the individual matrix eigenvalues~\cite{Shenker:1990uf}, and  their $\e^{-\hbar}$ effects. This ultimately  was connected to extended spacetime objects called D-branes~\cite{Polchinski:1995mt} upon which the string worksheets end,  crucial in the Second Superstring revolution that followed~\cite{Polchinski:1998rr} in the years leading up to the discovery of AdS/CFT. By then, matrix models had long been left behind, however. Even after interest in them was re-ignited in the early 2000s (see {\it e.g.,}~refs.~\cite{McGreevy:2003kb,Klebanov:2003km,McGreevy:2003ep,Martinec:2003ka,Alexandrov:2003nn,Schomerus:2003vv,Sen:2003iv,Klebanov:2003wg,Maldacena:2004sn}), their full non-perturbative content and understanding remained rather subtle.

The new context of JT gravity has forced a re-examination and upgrading of the non-perturbative understanding, in part because the black hole context requires a direct facing up to the question as to the location and precise nature of the  quantum gravity microstates.
The key point is to realize that random matrix models are primarily, as it says on the box they come in, models of random matrices. That they contain within them a limit where an expansion as sums over 2D Euclidean surfaces can be made ({\it \`a la} 't Hooft) is a handy feature,  but it should not be taken as their defining characteristic. In fact, as will be made clear shortly, understanding them fully as theories of matrices will be key to understanding {\it why} such a limit exists, and  illuminate the subtle manner by which the microstate structure of the gravity theory emerges. The series of works of refs.~\cite{Johnson:2019eik,Johnson:2020heh,Johnson:2020exp} was able to derive and solve an analogue of the non-linear differential equation of old that yields a fully non-perturbative  description of the double-scaled matrix model of JT gravity. 

\begin{figure}[t]
\centering
\includegraphics[width=0.43\textwidth]{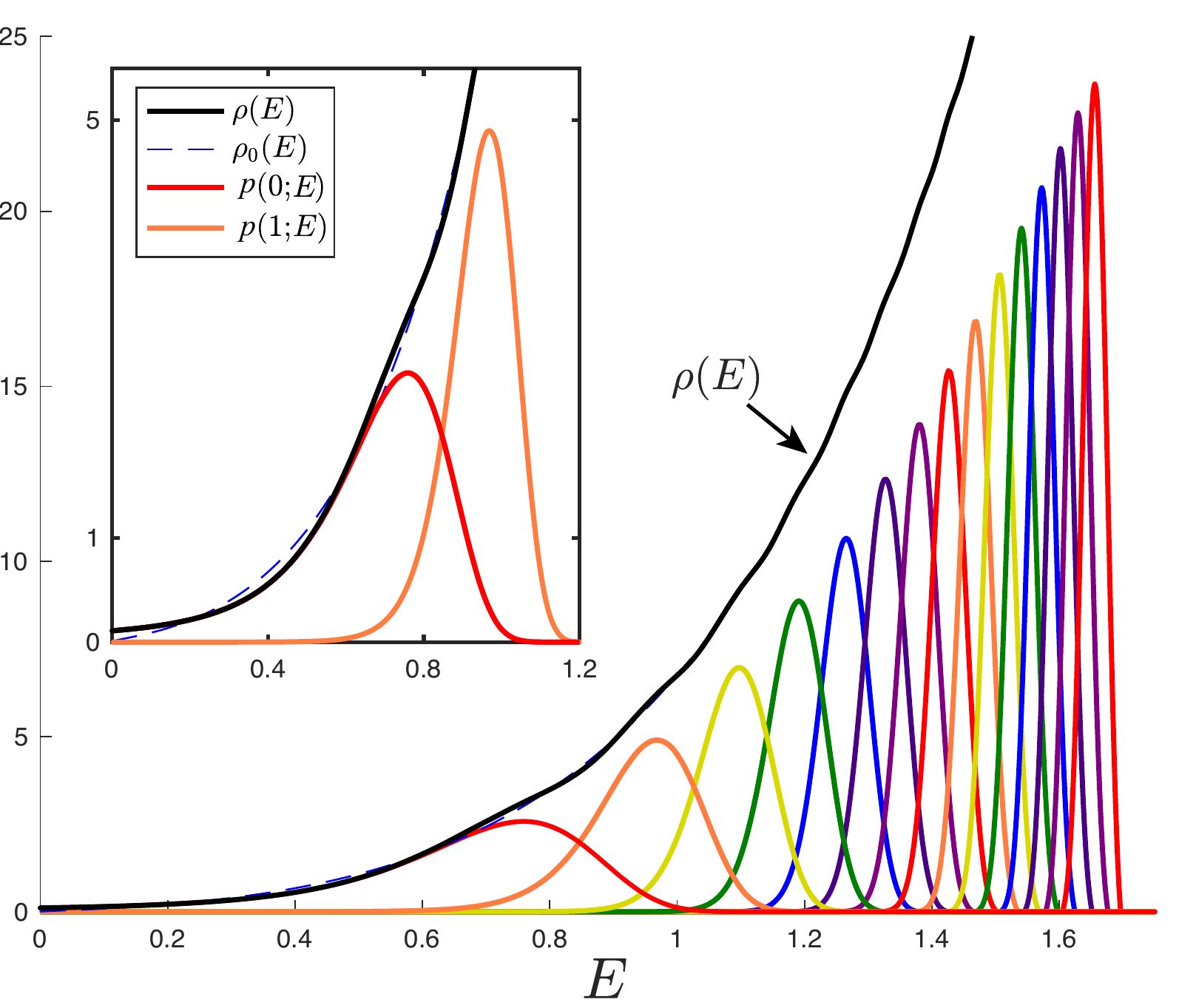}\hskip20pt
\includegraphics[width=0.5\textwidth]{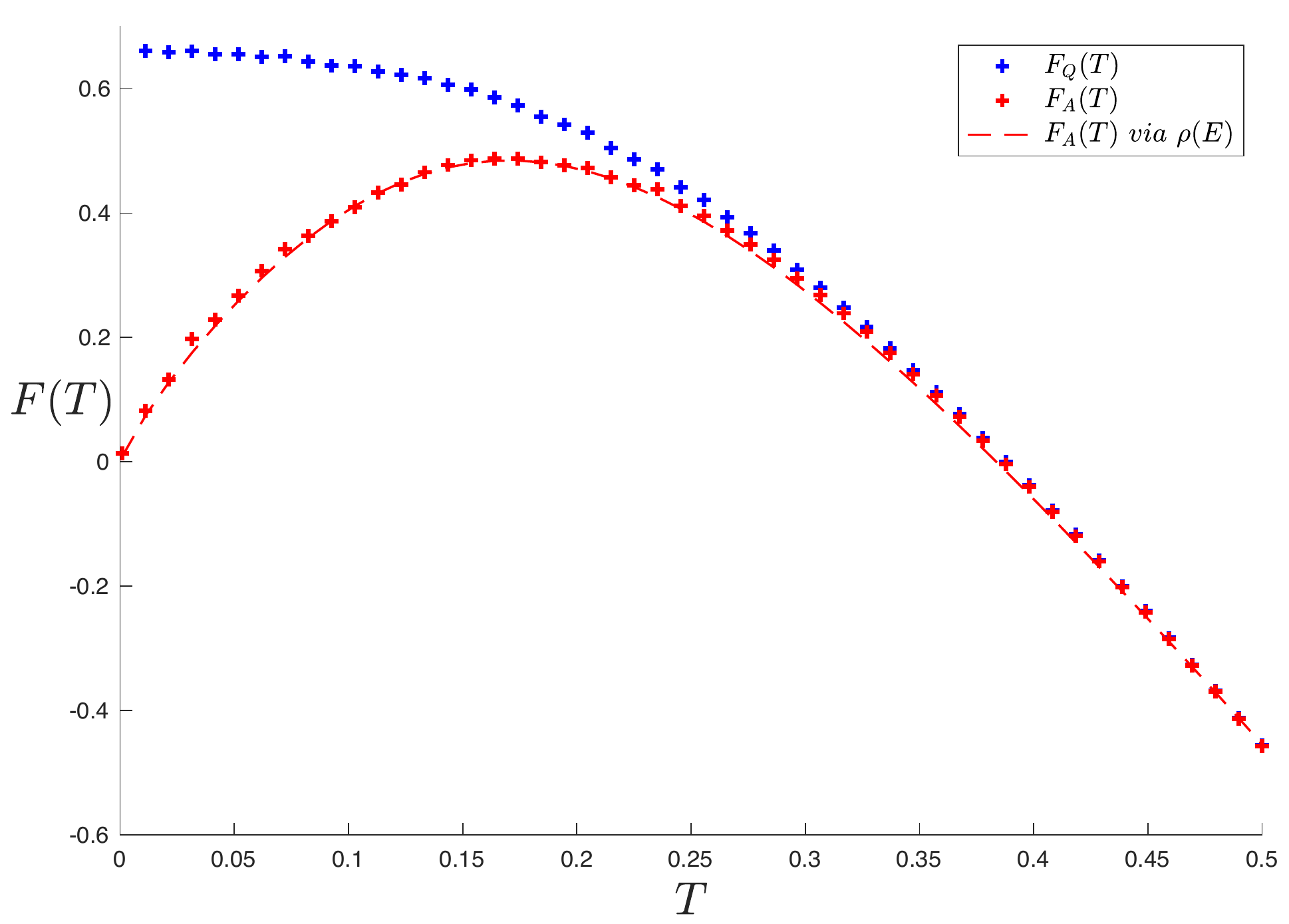}
\caption{\label{fig:JT-gravity-free-energy} \footnotesize {\bf Left:} 
Full spectral density $\rho(E)$ (solid black), leading part $\rho_0(E)$
 (blue dashed), and probability densities   of the first 15  levels of the  JT gravity
microstate spectrum. 
{\bf Right:} 
The quenched (blue) and annealed (red) free energies of JT gravity. 
 (Here, $\hbar{=}{\rm e}^{-S_0}{=}1$.)} 
\end{figure}

The resulting physics is identical to that of JT gravity perturbation theory of ref.~\cite{Saad:2019lba}, but the complete function $\rho(E)$ has much more besides.  See the black curve in figure~\ref{fig:JT-gravity-free-energy} (left). Again, it is a continuous function, although there is an interesting structure  of undulations, significant at low energy, and as $E$ grows, reducing in amplitude and increasing in frequency in such a way as to be negligible in the large $E$ (or large $S_0$) classical (perturbative) limit.  The nature of such undulations (at least a  good first draft of them) were already  deduced in the perturbative work of ref.~\cite{Saad:2019lba} by using semi-classical techniques, and they are indeed the first signs of the underlying discrete structure of the system. However, the full non-perturbative formulation gives complete information for all $E$, and additionally provides the  tools needed to fully appreciate what the matrix model is really doing.

{\it Enter Wigner}---Although the bumps in the spectral density were known to be connected to the discrete structure, just {\it how} they arise is important to know. For this, it is key to recognize that the spectral density $\rho(E)$ is really just a derived object from a more fundamental quantity that fully describes the matrix model physics. It is worth going back to where random matrix models first entered physics,  in the 1950s. Wigner~\cite{10.2307/1970079} was trying to characterize the properties of large complicated Hamiltonians, those of atomic nuclei. Although a particular nucleus {\it has} an answer for its spectrum, working it out is a difficult problem, and so the idea was to see if there are universal features of the class of Hamiltonians~$M$, drawn at random according to some probability $P(M)=\e^{-N{\rm Tr}[V(M)]}/{\tilde Z}$, where ${\tilde Z}$ is given in equation~(\ref{eq:partfun}). Wigner had in mind some Gaussian probability, {\it i.e.,} quadratic $V(M)$, but of course other polynomials are possible too. (Of course this connects with the random matrix models studied for gravity, where those higher powers of $M$ can lead to a description of surfaces in the 't Hooftian manner above, but  set that aside for now,  instead  following  the Wignerian path to see where it leads.)

The  problem of statistically characterizing the spectrum of the matrices is one of studying the probable locations of their eigenvalues $\lambda_i$ on the real line parameterized by coordinate~$\lambda$.  It all comes down to knowing the symmetric ``kernel'' $K(\lambda,\lambda^\prime)$,  a  function in terms of  which matrix model questions, phrased as probabilities, can be answered in terms of computing a  determinant built from it: 
The probability of finding $m$ eigenvalues at the positions $\lambda_i$ ($i{=}1,\cdots,m$) is given by the determinant of the $m\times m$ matrix   $K(\lambda_i,\lambda_j)$. 
%
%
Note that   $K(\lambda,\lambda){=}\rho(\lambda)$ is the basic spectral density function itself. 
Clearly there is a great deal more information in the kernel than just $\rho(\lambda)$.
%
 The double scaling limit turns out to be equivalent to focusing on the infinitesimal neighbourhood of the endpoint of the distribution of eigenvalues at large $N$, a region with critical (and universal) behaviour that is also  of interest to many other fields of statistical physics~\cite{Kazakov:1989bc,Dalley:1991zs,Bowick:1991ky,FORRESTER1993709,Meh2004,ForresterBook}. It has ``zoomed in'' coordinate~$E$, and  a scaling piece of the kernel $K(E,E^\prime)$  survives to allow the study properties of the spectrum.  
  In particular, one might ask  about the probability of there being {\it no} eigenvalues on an interval $(a,b)$ in the spectrum. As shown by Gaudin~\cite{Gaudin1961SurLL} in 1961, the answer can be phrased in terms of the   integral operator $\mathbf{K}|_{(a,b)}$ with  kernel $K(E,E^\prime)$, acting on the space $a{\leq}E{\leq}b$ according to $\int_a^bK(E,E^\prime)f(E^\prime)dE^\prime {=} g(E)$, for  some functions $f(E)$ and~$g(E)$.
The probability is in fact the Fredholm determinant~\cite{10.1007/BF02421317} ${\rm det}[\mathbb{I}-\mathbf{K}|_{(a,b)}]$.
Computations using this object  determines  the probability distribution of the lowest energy eigenvalue, the next lowest, and so on. For the Gaussian case, this results in the celebrated Tracy-Widom distribution~\cite{Tracy:1992rf}, a relatively new kind of universal distribution function  appearing in a wide range of mathematical and physics settings such as combinatorics, complex dynamical systems, trapped Coulomb gases, and stochastic growth processes~\cite{quanta:wolchover}. It is exciting therefore that quantum gravity calls for an analogue of it, and recent work~\cite{Johnson:2021zuo,Johnson:2022wsr} allows it to be computed for  JT gravity, building on the non-perturbative results already mentioned. The result is given in figure~\ref{fig:JT-gravity-free-energy} (left).

Now a new level of understanding appears. There in fact is an underlying discrete structure hidden in the continuous spectral density, by virtue of it being also  a  sum
$\rho(E){=}\sum_{n=0}^\infty p(n;E)$,
where a curve $p(n;E)$ (a coloured peak in the figure) is the probability density function for the $n$th energy level of a matrix in the ensemble.  In particular, the case of $n{=}0$ is the distribution of the ground state. It is clear that there's a notion of a most probable value of the ground state (the top of the peak), or the average ground state value (its mean). Similar things can be said about higher energy  levels. These are lessons Wigner would take away about how the spectrum of  a typical nucleus' Hamiltonian is characterized. This notion will be returned to at the end. 

The spacing of the peaks is   set by ${\rm e}^{-S_0}$,  as was  anticipated above on general grounds, but there is also an energy dependence: At higher energies the gaps reduce in size, and the peaks themselves become sharper. Once the energy regime 
 of the aforementioned leading Schwarzian computation  is reached, the peaks become a dense set of $\delta$--functions and the spectrum is effectively continuous,  reducing to the Schwarzian result~(\ref{eq:leading-relation}). It is clear now that the matrix model is built from discrete spectra of $\delta$--functions at some definite energy locations ${\cal E}_n$. There is an infinite set of such spectra with locations of the ${\cal E}_n$ distributed according to the $p(n;E)$, and so the smooth density function $\rho(E)$ of the matrix model is the result
%
\begin{wrapfigure}{r}{0.5\textwidth}
\centering
\vskip-0.5pt
\includegraphics[width=0.45\textwidth]{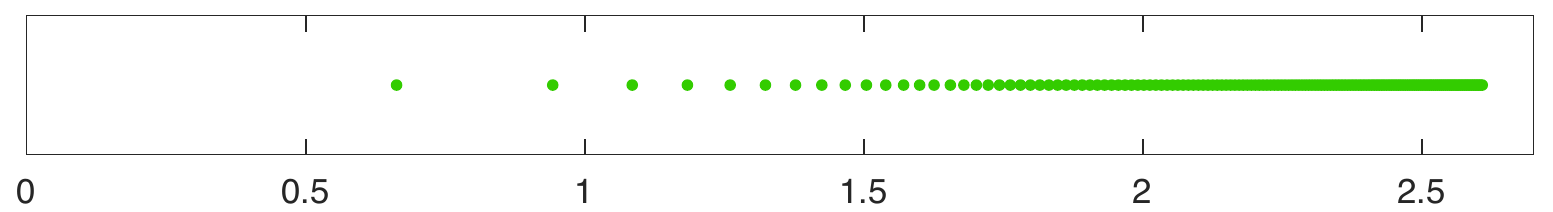}
\caption{\label{fig:JT-levels-150} \footnotesize  Mean values of the  first 150 microstates.}
\end{wrapfigure}
 of combining them.
At large $E$, asymptotically, they increasingly resemble each other, since the variance of the peaks reduces. In fact a simple approximate asymptotic formula for the discrete energies ${\cal E}_n$ at large level can be derived using the methods discussed in ref.~\cite{Johnson:2022wsr}.
 It is $\int_0^{{\cal E}_n}\rho_0(E) dE {=} n$, where $\rho_0(E)$ is the leading Schwarzian result. The level density is $dn/d{\cal E}_n$, which gives back $\rho_0({\cal E}_n)$. Taking~${\cal E}_n$ to be the mean of each peak to yield an ``average'' spectrum is instructive, and the first 150 are  plotted in figure~\ref{fig:JT-levels-150}, showing the journey to the continuum.

Of course, the exciting result here is that  finally we now have the  discrete quantum microstate structure called for by the old Bekenstein-Hawking result for the entropy all those decades ago! To see its properties in action it is useful to consider the full thermodynamics of the theory.  Attempting to compute in the manner done  when the physics is dominated by a single saddle rapidly runs into difficulty. Taking the full non-perturbative JT gravity $Z(\beta)$ computed by the matrix model  and forming $F(T)\stackrel{?}{=}{-}T\log  Z(T)$ gives an  answer that begins sensibly for large $T$ (the Schwarzian regime), where $S(T){=}{-}\partial F/\partial T{=}S_0+2\pi^2T+\cdots$, but eventually $F(T)$ reaches a maximum and  decreases, so the entropy    decreases to zero and then goes negative! Something has gone wrong. From the matrix model perspective this is the result of computing the ``annealed'' free energy $F_A(T){=}{-}T\log \langle Z(T)\rangle$, where the ensemble average $\langle\cdot\rangle$ over spectra is being performed in the matrix model. What should be computed is the ``quenched'' free energy $F_Q(T){=}{-}T\langle\log Z(T)\rangle$, as was first noticed in this context by Engelhardt, Fischetti, and Maloney~\cite{Engelhardt:2020qpv}. The language here comes from the theory of disordered systems (going back to Edwards and Anderson~\cite{Edwards_1975} in the context of spin glasses). Some partial results were obtained in various special limits~\cite{Engelhardt:2020qpv,Okuyama:2021pkf,Janssen:2021mek}, but no successful complete computation was done for random matrix models of gravity until refs.~\cite{Johnson:2021rsh,Johnson:2021zuo}, using the microstructure identified above.  See~figure~\ref{fig:JT-gravity-free-energy} (right), showing sensible thermodynamics all the way down to $T{=}0$. In fact, the full understanding~\cite{Johnson:2022wsr} of the picture maps beautifully on to  analogous computations in the disordered media literature.

Generically, the annealed free energy  of a system starts giving wrong answers at low temperatures because atypical configurations (here, spectra)  can make large contributions to the sum.  They simply add up in the average partition function, and then the log is taken, giving a free energy that is smaller than the true value. Generically, such atypical contributions  become more prevalent at low $T$, and here this  precisely matches the fact that the variance of the peaks grows at lower energy levels. On the other hand, the contributions of atypical  configurations wash out in the quenched free energy upon averaging the log of the partition sum. The final result for $F_Q(T)$  better represents the contribution of the most typical or average configurations of the system. Indeed, given the $p(n;E)$ it is possible to generate many sample spectra, compute the partition function for each, and explicitly compute the two free energies for the JT gravity matrix model. A check of the statements made here is that  the free energy of the single copy ``mean'' spectrum mentioned above (that was extracted from the individual peaks)  matches $F_Q(T)$ extremely well~\cite{Johnson:2022wsr}.

{\it Wigner meets 't Hooft near the horizon}---A central lesson, which is likely useful beyond this 2D setting, is that the Euclidean quantum gravity calculus alone cannot fully reveal the presence of microstates, although it can unearth rumours of the discreteness ({\it e.g.,} the finite Bekenstein-Hawking entropy). What was needed was not just the complete sum over all surfaces of all topologies, as given by 't Hooft's expansion of the matrix model, but a picture  that went well beyond reference to any notion of smooth Euclidean surfaces.

What is the meaning of this structure that lies beyond? What to make of these spectra that have been combined in a Wignerian way? Well, here interpretations differ, and it is the subject of ongoing debate. The currently popular interpretation is that the entire ensemble must be taken to define the quantum gravity. If one regards quantum gravity entirely as an Euclidean enterprise, then in a sense this is true. In fact the structures seen here show unequivocally  that 
the role of the entire ensemble is simply a clever means of building a smooth density function $\rho(E)$ by adding all the discrete spectra, whence it is expanded in $\hbar$ to yield the smooth surfaces at each order that are evidently needed to do Euclidean gravity. \\
\indent However,  there can be  another interpretation,  following from the path Wigner might have taken in studying the near-extremal black hole. One might seek, as  above,  the quantum gravity corrections   to the  spectrum beyond the leading large $S_0$ result, looking for some definite discrete structure signalled by the finite entropy $S_0$. Since  such black hole solutions can be embedded 
 into a complete theory of quantum gravity like string theory, there should be some definite answer, even if it is complicated to determine.  Wigner might have studied, just as for nuclei,  the ensemble of Hamiltonians of size $\e^{S_0}$, that  all have  leading spectrum  given by  $\rhoo(E)$ in equation~(\ref{eq:leading-relation}). Beyond using that, this is an {\it entirely Lorentzian pursuit}, seeking the likely properties of the definite Hamiltonian.  In this way, he would have arrived at a matrix model  of the form discussed here,  discovering a great deal of data about the likely form of the discrete spectrum. {\it No sums over Euclidean surfaces would be involved.} He might conclude that there is indeed some definite quantum mechanical spectrum, resembling the continuous Schwarzian at large~$E$, but  with the  typical discrete properties shown here. \\
\indent 
In short,  Wigner would meet 't Hooft coming the other way, but it is instead the 't~Hooft of the holographic principle!  In other words, while not able to compute the spectrum exactly,  the matrix model reveals the most likely key properties of the spectrum of the 1D holographic  dual to the nearly AdS$_2$  gravity describing the black hole under study. It is the  analogue of the 4D Yang-Mills theory dual of  AdS$_5$ gravity with which this      essay began.\\
\indent

{\it Acknowledgments}---CVJ thanks Ahmed Almheiri, Ibrahima Bah, Panos Betzios, Lorenz Eberhardt, Matthew Heydeman, Juan Maldacena, Henry Maxfield, Felipe Rosso,  Edgar Shaghoulian, Joaquin Turiaci, and Herman Verlinde for conversations, Edward Witten for a  lively discussion,  the  US Department of Energy for support under grant  \protect{DE-SC} 0011687, and  Amelia for her support and patience.

\newpage

\bibliographystyle{apsrev4-1}
\bibliography{Fredholm_super_JT_gravity1,Fredholm_super_JT_gravity2}

\end{document}